\documentclass[12pt]{article}

\author{Yu.~M.~Zinoviev${}^{a,b}$\thanks{yurii.zinoviev@ihep.ru}
\\[0.5cm]
\it{\small ${}^a$Institute for High Energy Physics of National
Research Center "Kurchatov Institute"}\\
\it{\small Protvino, Moscow Region, 142281, Russia} \\[0.3cm]
\it\small{${}^b$Moscow Institute of Physics and Technology (State
University),} \\
\it{\small Dolgoprudny, Moscow Region, 141701, Russia}} 

\title{On massive super(bi)gravity\\
in the constructive approach}

\date{}

\begin{document}

\maketitle

\begin{abstract}
In this paper we investigate the possible supersymmetric extensions
for the massive (bi)gravity theories in the lowest non-trivial order.
For this purpose we construct the cubic interaction vertices for
massive spin-2 and one or two massive spin-3/2 fields restricting
ourselves with the terms containing no more than one derivative so
that such models can be considered as the smooth deformations for the
usual (spontaneously broken) supergravity. Also we investigate all
possible limits where one of the fields becomes massless.
\end{abstract}

\thispagestyle{empty}
\newpage
\setcounter{page}{1}

\tableofcontents

\section{Introduction}

A few years ago an interesting solution of the longstanding problem of
massive deformation for the gravity theory was found
\cite{RGT10,RGT11}. The solution appeared to be surprisingly simple:
all that one has to do is to supplement the usual Lagrangian for
gravity with the potential (containing terms without derivatives only)
and to chooses a special form of this potential so that to avoid the
appearance of the so-called Boulware-Deser ghost \cite{BD72}. In some
sense this theory may be considered as a smooth deformation of the
usual gravity theory. Further on, the extension to the bigravity
(with one massless and one massive gravitons) was also found 
\cite{HRS11,HR11,HR11b}.

Taking into account a prominent role played by supersymmetry it is
natural to call for the supersymmetric extensions of these massive
gravity and bigravity theories. It is strange but till now there
appeared just a few papers devoted to this subject 
\cite{Mal13,Mal13a,MFG16,OT16}. In this work we begin an investigation
of possible supersymmetric generalizations for the massive (bi)gravity
theories using the constructive approach based on the gauge invariant
description for the massive spin-2 and spin-3/2 particles. It is clear
that in most cases a supersymmetry must be spontaneously broken and a
gravitino must be massive. So we start with the construction of the
cubic interaction vertices for the massive spin-2 and one or two
massive spin-3/2. By analogy with the case of massive (bi)gravity
itself, we consider the vertices containing no more then one
derivative so that such models can also be considered as the smooth
deformations of the usual (spontaneously broken) supergravity. Let us
briefly summarize the results of our work here.

\noindent
{\bf Massive spin-2 and one massive spin-3/2} The vertex exists for
any non-zero values for the graviton and gravitini masses. It has a
smooth limit where graviton mass goes to zero that corresponds to the
case of the spontaneously broken supergravity. At the same the limit
where gravitino mass goes to zero is singular and the reason is clear:
massless gravitino means unbroken supersymmetry and so graviton have
to be massless as well.

\noindent
{\bf Massive spin-2 and massive and massless spin-3/2} The solution
exists for the equal masses for the graviton and gravitino only. The
reason is again quite clear: massless gravitino implies unbroken
supersymmetry and so all the members of the same supermultiplet must
have equal masses. At the same time, unbroken supersymmetry means
that there must exists a massless spin-2 superpartner for the massless
spin-3/2 so that such model must be a part of some bigravity theory
similar to the ones considered in \cite{MFG16}. Moreover, this result
agrees with the general properties of the cubic vertices for one
massless and two massive fields \cite{Met07b} where cubic vertex with
different masses requires much higher number of derivatives than for
the case of equal masses. In particular, a cubic vertex for massless
spin-3/2 and massive spin-2 and massive spin-3/2 with different
masses does exist but requires as many as four derivatives so it is
trivially gauge invariant.

\noindent
{\bf Massive spin-2 and two massive spin-3/2 with different masses}
The vertex exists for any three non-zero masses. The limit where the
graviton becomes massless is possible for the equal masses
for the two gravitini only. Again it is in agreement with the general
properties of such cubic vertices \cite{Met07b}. Indeed the cubic
vertex for the massless spin-2 and two massive spin-3/2 with
different masses does exist but requires as many as four derivatives.
From the other hand the limit where one of the gravitini becomes
massless is possible for the equal masses for the graviton and
massive gravitini only in agreement with the results of the previous
case.

The paper is organized as follows. In Section 2 we describe version of
the constructive approach we use. Section 3 provides all necessary
kinematic information on the frame-like gauge invariant description for
massive spin-2 and massive spin-3/2. Section 4, 5 and 6 are devoted to
the three types of cubic vertices described above. Some technical
details are moved into Appendix.

\section{Constructive approach}

We follow the constructive approach where starting with the free
(quadratic) Lagrangian ${\cal L}_0$ which is invariant under the
non-homogeneous gauge transformations $\delta_0 \Phi$ one tries to
construct interacting theory perturbatively in the number of fields:
$$
{\cal L} = {\cal L}_0 + g {\cal L}_1 + \dots, \qquad
\delta \Phi = \delta_0 \Phi + g \delta_1 \Phi + {\cal O}(g^2)
$$
where ${\cal L}_1$ contains cubic terms, while $\delta_1 \Phi$ is
linear in fields and so on. In the first non-trivial approximation it
requires to solve the following relation:
\begin{equation}
\frac{\delta {\cal L}_1}{\delta \Phi} \delta_0 \Phi +
\frac{\delta {\cal L}_0}{\delta \Phi} \delta_1 \Phi = 0
\end{equation}
This solution can be found in two steps. First of all one can find the
cubic terms ${\cal L}_1$ such that their variations vanish on the free
mass shell:
$$
\left. \frac{\delta {\cal L}_1}{\delta \Phi} \delta_0 \Phi
\right|_{\frac{\delta {\cal L}_0}{\delta \Phi}=0} = 0
$$
and then returning to the equation (1) find the corresponding
corrections to the gauge transformations.

In the frame-like formalism one works with the pairs of physical and
auxiliary fields (we denote them schematically as $\Phi$ and $\Omega$)
and so in the honest first order formalism one has to solve the
following relation:
$$
\frac{\delta {\cal L}_1}{\delta \Phi} \delta_0 \Phi 
+ \frac{\delta {\cal L}_1}{\delta \Omega} \delta_0 \Omega
+ \frac{\delta {\cal L}_0}{\delta \Phi} \delta_1 \Phi 
+ \frac{\delta {\cal L}_0}{\delta \Omega} \delta_1 \Omega = 0
$$
Taking into account that the equations for the auxiliary fields are
purely algebraic ones, in supergravities the so-called 1 and 1/2 order
formalism is very often used:
$$
\left[ \frac{\delta {\cal L}_1}{\delta \Phi} \delta_0 \Phi 
+ \frac{\delta {\cal L}_0}{\delta \Phi} \delta_1 \Phi 
\right]_{\frac{\delta({\cal L}_0 + {\cal L}_1)}{\delta \Omega}=0}
 = 0
$$
where one takes into account variations of the physical fields only
but all the calculations are made up to the terms proportional to the
auxiliary field equations only. But such formalism requires to solve
the complete non-linear equations for the auxiliary fields and it can
be quite non-trivial task. There exists one more possibility that we
called a modified 1 and 1/2 order formalism \cite{Zin09,BSZ11}
\begin{equation}
\left[ \frac{\delta {\cal L}_1}{\delta \Phi} \delta_0 \Phi 
+ \frac{\delta {\cal L}_1}{\delta \Omega} \delta_0 \Omega
+ \frac{\delta {\cal L}_0}{\delta \Phi} \delta_1 \Phi
\right]_{\frac{\delta {\cal L}_0}{\delta \Omega}=0} = 0
\end{equation}
where all that one needs are the solutions for the free auxiliary
fields equations only. It is this formalism that we use in this work.

\section{Kinematics}

The most important ingredient of the constructive approach is the
presence of the gauge invariance already at the free level. That is
why the constructive approach is usually associated with the theories
of massless fields only. But the gauge invariant description of the
massive bosonic and fermionic fields \cite{Zin01,Met06,Zin08b,PV10},
which is possible due to the introduction of the appropriate set of
Stueckelberg fields, allows one to extend such approach to any systems
with massive and/or massless fields. In this section we provide all
necessary information on the frame-like gauge invariant description
for the massive spin-2 and spin-3/2 fields \cite{Zin08b}.

\noindent
{\bf Notations and conventions}
We work in the frame-like formalism where four dimensional flat
Minkowski space is described (in a coordinate free way) by the 
(non-dynamical) frame $e^a$, its inverse $\hat{e}_a$ and covariant
derivative D. We use the condensed notations for the products of these
forms:
$$
E^{ab} = e^a \wedge e^b, \qquad
E^{abc} = e^a \wedge e^b \wedge e^c, \qquad
E^{abcd} = e^a \wedge e^b \wedge e^c \wedge e^d
$$
and similarly for $\hat{e}_a$, while 
$$
D \wedge D = 0
$$
In what follows a wedge product sign $\wedge$ is omitted.

Completely antisymmetric products of $\gamma$-matrices are defined as
follows:
$$
\Gamma^{ab} = \frac{1}{2} \gamma^{[a} \gamma^{b]}, \qquad
\Gamma^{abc} = \frac{1}{3!} \gamma^{[a} \gamma^b \gamma^{c]}, \qquad
\Gamma^{abcd} = \frac{1}{4!} \gamma^{[a} \gamma^b \gamma^c \gamma^{d]}
$$
We use a Majorana representation for the $\gamma$-matrices where
$(\gamma^0\gamma^a)$ and $(\gamma^0\Gamma^{ab})$ are symmetric in
their spinor indices, while $(\gamma^0)$, $(\gamma^0\Gamma^{abc})$ and
$(\gamma^0\Gamma^{abcd})$ are antisymmetric.

\subsection{Massive spin-2}

The frame-like gauge invariant description for the massive spin-2
field \cite{Zin08b} requires three pairs of physical and auxiliary
fields: ($\Omega^{ab},f^a$), ($B^{ab},A$) and ($\pi^a,\sigma$), where
$\Omega^{ab}$, $f^a$ and $A$ are one-forms while $B^{ab}$, $\pi^a$ and
$\sigma$ --- zero-forms. In the notations explained above the free
Lagrangian has the form:
\begin{eqnarray}
{\cal L}_0 &=& \frac{1}{2} \hat{E}_{ab} \Omega^a{}_c \Omega^{bc} -
\frac{1}{2} \hat{E}_{abc} \Omega^{ab} D f^c + \frac{1}{2} B_{ab}
B^{ab} \nonumber \\
 && - \hat{E}_{ab} B^{ab} D A - \frac{1}{3} \pi_a \pi^a + 
\frac{2}{3} \hat{e}_a \pi^a D \sigma \nonumber \\
 && + m \hat{E}_{ab} \Omega^{ab} A + m \hat{e}_a B^{ab} f_b - 2m
\hat{e}_a \pi^a A \nonumber \\
 && + \frac{m^2}{2} \hat{E}_{ab} f^a f^b - m^2 \hat{e}_a f^a \sigma +
\frac{2m^2}{3} \sigma^2
\end{eqnarray}
Its structure follows the general pattern for the gauge invariant
Lagrangians for the massive fields. Namely, the first two lines are
just the sum of the kinetic terms for the massless spin-2, spin-1 and
spin-0 fields, the last line is the sum of all possible mass-like
terms, while the third line contains cross-terms gluing all the fields
together. The main requirement determining this structure is that the
Lagrangian must still be invariant under the all (appropriately
modified) gauge transformations of the initial massless fields. Indeed
it is straightforward to check that this Lagrangian is invariant under
the following gauge transformations:
\begin{eqnarray}
\delta \Omega^{ab} &=& D \eta^{ab} - \frac{m^2}{2} e^{[a} \xi^{b]}
\nonumber \\
\delta f^a &=& D \xi^a - e_b \eta^{ab} + m e^a \xi \nonumber \\
\delta B^{ab} &=& - m \eta^{ab}, \qquad
\delta A = D \xi + \frac{m}{2} e_a \xi^a \\
\delta \pi^a &=& - \frac{3m^2}{2} \xi^a, \qquad
\delta \sigma = 3m \xi \nonumber
\end{eqnarray}
One of the nice features of the frame-like formalism is that for each
field (both physical and auxiliary ones) one can construct the
corresponding gauge invariant object. For the case at hands we obtain:
\begin{eqnarray}
{\cal R}^{ab} &=& D \Omega^{ab} + \frac{m}{2} E^{[a}{}_c B^{b]c} -
\frac{m^2}{2} e^{[a} f^{b]} + \frac{m^2}{3} E^{ab} \sigma \nonumber \\
{\cal T}^a &=& D f^a - e_b \Omega^{ab} + m e^a A \nonumber \\
{\cal B}^{ab} &=& D B^{ab} + m \Omega^{ab} - \frac{m}{3}
e^{[a} \pi^{b]} \nonumber \\
{\cal A} &=& D A - \frac{1}{2} E_{ab} B^{ab} + \frac{m}{2} e_a
f^a \\
\Pi^a &=& D \pi^a - \frac{3m}{2} e_b B^{ab} + \frac{3m^2}{2} f^a -
\frac{m^2}{2} e^a \sigma \nonumber \\
\Sigma &=& D \sigma - e_a \pi^a - 3m A \nonumber
\end{eqnarray}
In what follows we call these objects curvatures though 
${\cal R}^{ab}$, ${\cal T}^a$ and ${\cal A}$ are the two-forms, while
${\cal B}^{ab}$, $\Pi^a$ and $\Sigma$ --- one-forms.

As we have explained in the previous section all the calculations
are made up to the terms proportional to the free auxiliary fields
equations. So in what follows "on-shell" means:
\begin{equation}
{\cal T}^a \approx 0, \qquad {\cal A} \approx 0, \qquad \Sigma \approx
0
\end{equation}
This in turn provides us with a number of algebraic and differential
identities for the curvatures that do not vanish on-shell:
\begin{equation}
e_b {\cal R}^{ab} \approx 0, \qquad E_{ab} {\cal B}^{ab} \approx 0,
\qquad e_a \Pi^a \approx 0
\end{equation}
\begin{equation}
D {\cal R}^{ab} = \frac{m}{2} E^{[a}{}_c {\cal B}^{b]c}, \qquad
D {\cal B}^{ab} = m {\cal R}^{ab} + \frac{m}{3} e^{[a} \Pi^{b]},
\qquad D \Pi^a = \frac{3m}{2} e_b {\cal B}^{ab}
\end{equation}
One more useful fact is that variations of the free Lagrangian under
any transformations for the physical fields can be conveniently
expressed in terms of these curvatures:
\begin{equation}
\delta {\cal L}_0 = - \frac{1}{2} \hat{E}_{abc} {\cal R}^{ab} \delta
f^c + \hat{E}_{ab} {\cal B}^{ab} \delta A - \frac{2}{3} \hat{e}_a
\Pi^a \delta \sigma
\end{equation}

\subsection{Massive spin-3/2}

For the frame-like gauge invariant description of the massive spin-3/2
field we use a one-form $\Phi$ and a zero-form $\phi$ (both of them
are physical) with the Lagrangian
\begin{eqnarray}
{\cal L}_0 &=& - \frac{i}{2} \hat{E}_{abc} \bar{\Phi} \Gamma^{abc} D
\Phi + \frac{i}{2} \hat{e}_a \bar{\phi} \gamma^a D \phi \nonumber \\
 && - \frac{3m_1}{2} \hat{E}_{ab} \bar{\Phi} \Gamma^{ab} \Phi +
3im_1 \hat{e}_a \bar{\Phi} \gamma^a \phi - m_1 \bar{\phi} \phi
\end{eqnarray}
where the first line is just the sum of the kinetic terms for the
massless spin-3/2 and spin-1/2 fields, while the second line contains
mass-like and cross terms. This Lagrangian is invariant under the
following gauge transformations:
\begin{equation}
\delta \Phi = D \zeta + \frac{im_1}{2} e_a \gamma^a \zeta, \qquad
\delta \phi = 3m_1 \zeta
\end{equation}
As in the bosonic case we can construct two gauge invariant objects
(curvatures):
\begin{eqnarray}
{\cal F} &=& D \Phi + \frac{im_1}{2} e_a \gamma^a \Phi + 
\frac{m_1}{12} E_{ab} \Gamma^{ab} \phi \nonumber \\
{\cal C} &=& D \phi - 3m_1 \Phi + \frac{im_1}{2} e_a \gamma^a \phi
\end{eqnarray}
where ${\cal F}$ is a two-form, while ${\cal C}$ --- one-form. These
curvatures satisfy the following differential identities:
\begin{eqnarray}
D {\cal F} &=& - \frac{im_1}{2} e_a \gamma^a {\cal F}
+ \frac{m_1}{12} E_{ab} \Gamma^{ab} {\cal C} \nonumber \\
D {\cal C} &=& - 3m_1 {\cal F} - \frac{im_1}{2} e_a \gamma^a {\cal C}
\end{eqnarray}
Also the variations of the free Lagrangian under any
transformations of the fields $\Phi$ and $\phi$ can be conveniently
expressed in terms of these curvatures:
\begin{equation}
\delta {\cal L}_0 = - i \hat{E}_{abc} \bar{\cal F} \Gamma^{abc}
\delta \Phi - i \hat{e}_a \bar{\cal C} \gamma^a \delta \phi
\end{equation}

For the second spin 3/2 we use the same formulas but with the fields 
$\Psi$ and $\psi$, mass $m_2$ and the gauge invariant curvatures 
${\cal H}$ and ${\cal D}$. Note also that in the massless limit 
$m_2 = 0$ the spinor field $\psi$ decouples and we obtain simply
\begin{equation}
{\cal L}_0 = - \frac{i}{2} \hat{E}_{abc} \bar{\Psi} \Gamma^{abc} D
\Psi, \qquad \delta \Psi = D \zeta
\end{equation}

\section{Massive spin-2 and one massive spin-3/2} \label{case1}

In this section we consider a cubic vertex for the massive spin-2 and
one massive spin-3/2 with different masses. We follow top-down
approach in the number of derivatives. Namely, we begin with the most
general non-trivial (i.e. such that do not vanish and are not
equivalent on-shell) terms with one derivative and require that all
variations with the highest number of derivatives can be compensated
by the appropriate corrections to the gauge transformations. Then we
add terms without derivatives and try to achieve complete invariance
introducing additional corrections if necessary.

\noindent
{\bf Terms with one derivative.} Complete analysis of these terms is
given in the Appendix while here we provide the result only:
\begin{eqnarray}
{\cal L}_{11} &=& ic_1 \hat{E}_{abc} \Omega^{ab} \bar{\Phi} \gamma^c
\Phi + ic_2 \hat{E}_{abcd} \bar{\cal F} \Gamma^{abc} \Phi f^d
\nonumber \\
 && + ic_3 \hat{e}_a \Omega^{bc} \bar{\phi} \Gamma^{abc}
\phi + ic_4 \hat{E}_{ab} \bar{\cal C} \gamma^a \phi f^b \nonumber \\
 && + ic_5 \hat{e}_a B^{ab} \bar{\Phi} \gamma^b \phi + ic_6
\hat{e}_a B^{bc} \bar{\Phi} \Gamma^{abc} \phi \nonumber \\
 && + c_7 \hat{e}_a \bar{\Phi} \phi \pi^a + c_8 \hat{e}_a
\bar{\Phi} \Gamma^{ab} \phi \pi^b 
\end{eqnarray}
The four lines in this expression correspond to the elementary
subvertices (2,3/2,3/2), (2,1/2,1/2), (1,3/2,1/2) and (0,3/2,1/2)
respectively. The possibility to compensate for the variations with
the highest number of derivatives arises when
$$
2c_1 = 3c_2, \qquad c_4 = - 4c_3, \qquad
c_5 = - 2c_6, \qquad c_8 = - c_7
$$
while the necessary corrections to the gauge transformations have the
form:
\begin{eqnarray}
\delta_1 f^a &=& - 4ic_1 \bar{\Phi} \gamma^a \zeta, \qquad
\delta_1 A = ic_6 \bar{\phi} e_a \gamma^a \zeta, \qquad
\delta_1 \sigma = - \frac{3c_7}{2} \bar{\phi} \zeta \nonumber \\
\delta_1 \Phi &=& - \frac{c_1}{3} \Gamma^{ab} \Omega^{ab} \zeta
+ \frac{c_1}{3} \Gamma^{ab} \eta^{ab} \Phi \\
\delta_1 \phi &=& - 2c_3 \Gamma^{ab} \eta^{ab} \phi
- c_6 \Gamma^{ab} B^{ab} \zeta + ic_7 \gamma^a \pi^a \zeta \nonumber
\end{eqnarray}

\noindent
{\bf Terms without derivatives} Now we proceed and add the most
general cubic terms without derivatives: 
\begin{eqnarray}
{\cal L}_{10} &=& d_1 \hat{E}_{abc} \bar{\Phi} \Gamma^{ab} \Phi f^c
 + id_2 \hat{E}_{ab} \bar{\Phi} \gamma^a \phi f^b 
 + id_3 \hat{E}_{ab} \bar{\Phi} \Gamma^{abc} \phi f^c 
 + d_4 \hat{e}_a \bar{\phi} \phi f^a \nonumber \\
 && + d_5 \hat{E}_{ab} \bar{\Phi} \Gamma^{ab} \phi A
 + d_6 \hat{E}_{ab} \bar{\Phi} \Gamma^{ab} \Phi \sigma
 + id_7 \hat{e}_a \bar{\Phi} \gamma^a \phi \sigma 
 + d_8 \bar{\phi} \phi \sigma
\end{eqnarray}
This in turn requires to
introduce additional corrections to the gauge transformations for the
fermions:
\begin{eqnarray}
\delta_1 \Phi &=& im_1c_2 \xi^a \gamma^a \Phi + \frac{mc_6}{3} e^a
\xi^a \phi + \frac{d_2}{6} e_a \Gamma^{ab} \xi_b \phi \nonumber \\
 && - im_1c_2 \gamma^a f^a \zeta - mc_2 A \zeta -
\frac{id_6}{3} e^a \gamma^a \zeta \sigma + mc_2 \Phi \xi \\
\delta_1 \phi &=& im_1c_4 \gamma^a \xi^a \phi + d_7 \zeta
\sigma + 3mc_4 \phi \xi \nonumber 
\end{eqnarray}
The complete gauge invariance (in the linear approximation) leads to
the number of relations on the parameters. Their solution looks like:
$$
c_3 = \frac{m^2-3m_1{}^2}{18m_1{}^2} c_1, \qquad
c_6 = - \frac{m}{3m_1} c_1, \qquad
c_7 = \frac{4}{9} c_1
$$
$$
d_1 = m_1c_1, \qquad
d_2 = \frac{2(m^2-2m_1{}^2)}{3m_1} c_1, \qquad
d_3 = - \frac{m_1}{3}c_1, \qquad
d_4 = - 2m_1c_3
$$
$$
d_5 = 0, \qquad
d_6 = - \frac{3m_1}{2} c_7, \qquad
d_7 = - d_2, \qquad
d_8 = \frac{8m_1}{3} c_3
$$
Thus all the parameters are expressed in terms of the main one $c_1$.

\noindent{\bf Algebraic structure.} All our fields are gauge or
Stueckelberg ones with the non-homogeneous gauge transformations. So
even in this linear approximation we can consider the commutators of
the gauge transformations in the lowest order and this gives quite
important information on the algebraic structure that stays behind
such a model and also provides an independent check for our
calculations. For the bosonic fields we can take the commutators of
the two supertransformations. This gives:
$$
[\delta_1, \delta_2] f^a = D \tilde{\xi}^a - \tilde{\eta}^{ab} e_b,
\qquad [\delta_1, \delta_2] A = \frac{m}{2} e_a \tilde{\xi}^a, \qquad
[\delta_1, \delta_2] \sigma = 0
$$
\begin{equation}
\tilde{\xi}^a = 4ic_1 (\bar{\zeta}_a \gamma^a \zeta_1), \qquad
\tilde{\eta}^{ab} = 4m_1c_1 (\bar{\zeta}_2 \Gamma^{ab} \zeta_1),
\qquad \tilde{\xi} = 0
\end{equation}
At the same time for the fermionic fields we have non-trivial
commutators for the bosonic and supertransformations:
$$
[ \delta_B, \delta_\zeta] \Phi = (D + \frac{im}{2} e_a \gamma^a)
\tilde{\zeta}, \qquad [ \delta_B, \delta_\zeta] \phi = 3m_1
\tilde{\zeta}
$$
\begin{equation}
\tilde{\zeta} = - \frac{c_1}{3} (\Gamma^{ab} \eta^{ab} \zeta)
 - \frac{2im_1c_1}{3} (\gamma^a \xi^a \zeta) 
 - \frac{2mc_1}{3} (\xi \zeta)
\end{equation}

\noindent
{\bf Massless limits.} From the solution for the parameters $c$, $d$
given above one can see that the limit $m_1 \to 0$ where gravitino
becomes massless is singular. It is quite natural because massless
gravitino means unbroken supersymmetry and in this case graviton
should also be massless. On the other hand, nothing prevent us to
consider the limit $m \to 0$ where graviton becomes massless  and this
corresponds to the case of spontaneously broken supergravity. The
cubic vertex and the corrections to the gauge transformations have the
form (compare \cite{BSZ14}):
\begin{eqnarray}
{\cal L}_1 &=& ic_1 \hat{E}_{abc} \Omega^{ab} \bar{\Phi} \gamma^c
\Phi + \frac{2ic_1}{3} \hat{E}_{abcd} \bar{\cal F} \Gamma^{abc} \Phi
f^d - \frac{ic_1}{6} \hat{e}_a \Omega^{bc} \bar{\phi} \Gamma^{abc}
\phi + \frac{2ic_1}{3} \hat{E}_{ab} \bar{\cal C} \gamma^a \phi f^b
\nonumber \\
 && + m_1c_1 \hat{E}_{abc} \bar{\Phi} \Gamma^{ab} \Phi f^c
 - \frac{im_1c_1}{3} \hat{E}_{ab} \bar{\Phi} (4 \gamma^a f^b 
 + \Gamma^{abc} f^c ) \phi  + \frac{m_1c_1}{3} \hat{e}_a \bar{\phi}
\phi f^a
\end{eqnarray}

\begin{eqnarray}
\delta_1 f^a &=& - 4ic_1 \bar{\Phi} \gamma^a \zeta \nonumber \\
\delta_1 \Phi &=& \frac{c_1}{3} \Gamma^{ab} (\eta^{ab} \Phi -
\Omega^{ab} \zeta) - \frac{im_1c_1}{6} (\xi^a \gamma^a \Phi - \gamma^a
f^a \zeta) - \frac{2m_1c_1}{9} e_a \Gamma^{ab} \xi_b \phi \\
\delta_1 \phi &=& \frac{c_1}{3} \Gamma^{ab} \eta^{ab} \phi + 
\frac{2im_1c_1}{3} \gamma^a \xi^a \phi \nonumber
\end{eqnarray}

\section{Massive spin-2 and massive and massless spin-3/2}
\label{case2}

In this section we consider the cubic vertex for the massive spin-2
and one massive and one massless spin-3/2. It is a special limit of
the general vertex that we consider in the next section but it is
worth to be considered separately.

\noindent
{\bf Terms with one derivative.} As in the previous case we begin with
the most general non-trivial terms with one derivative. The analysis
of the possible terms goes similarly to the previous case so we give
here the final result only:
\begin{eqnarray}
{\cal L}_{11} &=& 3ic_1 \hat{E}_{abc} \Omega^{ab} \bar{\Phi} \gamma^c
\Psi + ic_1 \hat{E}_{abcd} [ \bar{\cal F} \Gamma^{abc} \Psi f^d +
\bar{\cal H} \Gamma^{abc} \Phi ] f^d \nonumber \\
 && + c_3 \hat{E}_{ab} [ 2 B^{ab} \bar{\Phi} \Psi + B^{cd} \bar{\Phi}
\Gamma^{abcd} \Psi ] \nonumber \\
 && + ic_5 \hat{e}_a \bar{\Psi} [ 2 B^{ab} \gamma^c - B^{bc}
\Gamma^{abc} ] \phi \nonumber \\
 && + c_8 \hat{e}_a \bar{\Psi} [ \pi^a - \Gamma^{ab} \pi^b ] \phi
\end{eqnarray}
The four lines in this expression correspond to the subvertices
(2,3/2,3/2), (1,3/2,3/2), (1,3/2,1/2) and (0,3/2,1/2) respectively.
The corrections to the gauge transformations for the bosonic fields
which are necessary to compensate for the variations with the highest
number of derivatives look as follows:
\begin{eqnarray}
\delta f^a &=& - 6ic_1 [ \bar{\Psi} \gamma^a \zeta_1 + \bar{\Phi}
\gamma^a \zeta_2 ] \nonumber \\
\delta A &=& 2c_3 [ \bar{\Psi} \zeta_1 - \bar{\Phi} \zeta_2 ] -
ic_5 \bar{\phi} \gamma^a e_a \zeta_2 \\
\delta \sigma &=& - \frac{3c_8}{2} \bar{\phi} \zeta_2 \nonumber
\end{eqnarray}
while the corresponding corrections for the fermionic fields have the
form:
\begin{eqnarray}
\delta \Phi &=& \frac{c_1}{2} [ \Gamma^{ab} \eta^{ab} \Psi -
\Gamma^{ab} \Omega^{ab} \zeta_2] - \frac{ic_3}{6} [ 2 B^{ab} e_a
\gamma_b + \Gamma^{abc} B_{ab} e_c ] \zeta_2 \nonumber \\
\delta \phi &=& c_5 \Gamma^{ab} B^{ab} \zeta_2 + ic_8 \gamma^a \pi^a
\zeta_2 \\
\delta \Psi &=& \frac{c_1}{2} [ \Gamma^{ab} \eta^{ab} \Phi -
\Gamma^{ab} \Omega^{ab} \zeta_1 ] + \frac{ic_3}{6} [ 2 B^{ab} e_a
\gamma_b - \Gamma^{abc} B_{ab} e_c ] \zeta_1 \nonumber
\end{eqnarray}

\noindent
{\bf Terms without derivatives.} Now we proceed and add the most
general cubic terms without derivatives\footnote{The coefficients are
chosen so that they correspond to the similar coefficients in
the general case considered in the next section}: 
\begin{eqnarray}
{\cal L}_{10} &=& d_1 \hat{E}_{abc} \bar{\Phi} \Gamma^{ab} \Psi f^c +
d_2 \hat{E}_{abc} \bar{\Phi} \Gamma^{abcd} \Psi f^d + id_4 
\hat{E}_{ab} \bar{\Psi} \gamma^a \phi f^b + id_6 \hat{E}_{ab}
\bar{\Psi} \Gamma^{abc} \phi f^c \nonumber \\
 && + id_9 \hat{E}_{abc} \bar{\Phi} \Gamma^{abc} \Psi A + d_{13}
\hat{E}_{ab} \bar{\Phi} \Gamma^{ab} \Psi \sigma + id_{15} \hat{e}_a
\bar{\Psi} \gamma^a \phi \sigma
\end{eqnarray}
First of all to cancel all
the remaining variations we have to put $m_1 = m$, i.e. solution
exists for the equal masses for the graviton and gravitino only. The
reason is quite clear: massless gravitino implies unbroken
supersymmetry and so all the members of the same supermultiplet must
have equal masses. At the same time, unbroken supersymmetry means
that there must exists a massless spin-2 superpartner for the massless
spin-3/2 so that such model must be a part of some bigravity theory
similar to the ones considered in \cite{MFG16}. Moreover, this result
agrees with the general properties of the cubic vertices for one
massless and two massive fields \cite{Met07b} where cubic vertex with
different masses requires much higher number of derivatives than for
the case of equal masses. In particular, a cubic vertex for massless
spin-3/2 and massive spin-2 and massive spin-3/2 with different
masses does exist but requires as many as four derivatives so it is
trivially gauge invariant.

As in the previous case all the parameters are
expressed in terms of the main one $c_1$:
$$
c_3 = c_5 = \frac{3c_1}{2}, \qquad c_8 = 2c_1
$$
$$
d_1 = \frac{3mc_1}{2}, \qquad d_2 = \frac{mc_1}{2}, \qquad
d_4 = mc_1, \qquad d_6 = - \frac{mc_1}{2}
$$
$$
d_9 = 2mc_1, \qquad d_{13} = - 3mc_1, \qquad d_{15} = mc_1
$$
with the additional corrections for the fermionic fields:
\begin{eqnarray}
\delta \Phi &=& imc_1 \gamma^a \xi^a \Psi - mc_1 \Psi \xi - imc_1
\gamma^a f^a \zeta_2 + mc_1 A \zeta_2 + \frac{imc_1}{2} e_a \gamma^a
\sigma \zeta_2 \nonumber \\
\delta \phi &=& mc_1 \sigma \zeta_2    \\
\delta \Psi &=& - \frac{mc_1}{6} [ 3 e_a \xi^a - \Gamma^{ab} e_a
\xi_b] \phi + 3mc_1 \Phi \xi - 3mc_1 A \zeta_1 + \frac{imc_1}{2} e_a
\gamma^a \sigma \zeta_1 \nonumber
\end{eqnarray}

\noindent
{\bf Algebraic structure.} Again it is instructive to consider the
commutators of the gauge transformations in the lowest order. For the
bosonic fields the commutators of the two supertransformations have
the form:
$$
[\delta_1, \delta_2] f^a = D \tilde{\xi}^a - \tilde{\eta}^{ab} e_b + m
e^a \tilde{\xi}, \qquad [\delta_1, \delta_2] A = D \tilde{\xi}
+\frac{m}{2} e_a \tilde{\xi}^a, \qquad [\delta_1, \delta_2] \sigma =
3m \tilde{\xi}
$$
\begin{equation}
\tilde{\xi}^a = 6ic_1 (\bar{\zeta}_a \gamma^a \zeta_1), \qquad
\tilde{\eta}^{ab} = 3m_1c_1 (\bar{\zeta}_2 \Gamma^{ab} \zeta_1),
\qquad \tilde{\xi} = - 3c_1 (\bar{\zeta}_2 \zeta_1)
\end{equation}
while the commutators of the bosonic and supertransformations look
like:
$$
[ \delta_B, \delta_\zeta] \Phi = (D + \frac{im}{2} e_a \gamma^a)
\tilde{\zeta}_1, \qquad [ \delta_B, \delta_\zeta] \phi = 3m_1
\tilde{\zeta}_1
$$
\begin{equation}
\tilde{\zeta}_1 = - \frac{c_1}{2} (\Gamma^{ab} \eta^{ab} \zeta_2)
 - im_1c_1 (\gamma^a \xi^a \zeta_2) 
 + m_1c_1 (\xi \zeta_2)
\end{equation}
\begin{equation}
[ \delta_B, \delta_\zeta] \Psi = D \tilde{\zeta}_2, \qquad
\tilde{\zeta}_2 = - \frac{c_1}{3} (\Gamma^{ab} \eta^{ab} \zeta_1)
 - 3m_1c_1 (\xi \zeta_1)
\end{equation}

\section{Massive spin-2 and two spin-3/2 with different masses}

At last we consider the most general case --- massive spin-2 and two
massive spin-3/2 with different masses. 

\noindent
{\bf Terms with one derivative.} In this case we have quite a lot of
possible terms with one derivative:
\begin{eqnarray}
{\cal L}_{11} &=& 3ic_1 \hat{E}_{abc} \Omega^{ab} \bar{\Phi} \gamma^c
\Psi + ic_1 \hat{E}_{abcd} \bar{\cal F} \Gamma^{abc} \Psi f^d
+ ic_1 \hat{E}_{abcd} \bar{\cal H} \Gamma^{abc} \Phi f^d \nonumber \\
 && + 4ic_2 \hat{E}_{ab} \bar{\cal C} \gamma^a \psi f^b +
4ic_2 \hat{E}_{ab} \bar{\cal D} \gamma^a \phi f^b + ic_2 \bar{E}_{ab}
\bar{\cal C} \Gamma^{abc} \psi f^c + ic_2 \hat{E}_{ab} \bar{\cal D}
\Gamma^{abc} \phi f^c \nonumber \\
 && + 2c_3 \hat{E}_{ab} B^{ab} \bar{\Phi} \Psi 
 + c_3 \hat{E}_{ab} B^{cd} \bar{\Phi} \Gamma^{abcd} \Psi \nonumber \\
 && + 2ic_4 \hat{e}_a B^{ab} \bar{\Phi} \gamma^b \psi
 - ic_4 \hat{e}_a B^{bc} \bar{\Phi} \Gamma^{abc} \psi 
 + 2ic_5 \hat{e}_a B^{ab} \bar{\Psi} \gamma^b \phi
 - ic_5 \hat{e}_a B^{bc} \bar{\Psi} \Gamma^{abc} \phi \nonumber \\
 && + c_6 B^{ab} \bar{\phi} \Gamma^{ab} \psi \nonumber \\
 && + c_7 \hat{e}_a \bar{\Phi} \psi \pi^a - c_7 \hat{e}_a
\bar{\Phi} \Gamma^{ab} \psi \pi^b + c_8 \hat{e}_a \bar{\Psi} \phi
\pi^a - c_8 \hat{e}_a \bar{\Psi} \Gamma^{ab} \phi \pi^b 
\end{eqnarray}
where separate lines correspond to the subvertices (2,3/2,3/2),
(2,12,1/2), (1,3/2,3/2), (1,3/2,1/2), (1,1/2,1/2) and (0,3/2,1/2).
The required corrections to the gauge transformations for the bosonic
fields look like:
\begin{eqnarray}
\delta f^a &=& - 6ic_1 \bar{\Psi} \gamma^a \zeta_1 - 6ic_1 \bar{\Phi}
\gamma^a \zeta_2 \nonumber \\
\delta A &=& 2c_3 \bar{\Psi} \zeta_1 - 2c_3 \bar{\Phi} \zeta_2
 - ic_4 \bar{\psi} \gamma^a e_a \zeta_1 - ic_5 \bar{\phi} \gamma^a e_a
\zeta_2 \\
\delta \sigma &=& - \frac{3c_7}{2} \bar{\psi} \zeta_1 - \frac{3c_8}{2}
\bar{\phi} \zeta_2 \nonumber
\end{eqnarray}
while the corresponding corrections for the fermions have the form:
\begin{eqnarray}
\delta \Phi &=& \frac{c_1}{2} \Gamma^{ab} \eta^{ab} \Psi 
- \frac{c_1}{2} \Gamma^{ab} \Omega^{ab} \zeta_2 
 - \frac{ic_3}{3} B^{ab} e_a \gamma_b \zeta_2 + \frac{ic_3}{6}
\Gamma^{abc} B_{ab} \gamma_c \zeta_2 \nonumber \\
\delta \phi &=& 2c_2 \Gamma^{ab} \eta^{ab} \psi + c_5 \Gamma^{ab}
B^{ab} \zeta_2 + ic_8 \gamma^a \pi^a \zeta_2
\end{eqnarray}
\begin{eqnarray}
\delta \Psi &=& \frac{c_1}{2} \Gamma^{ab} \eta^{ab} \Phi 
 - \frac{c_1}{2} \Gamma^{ab} \Omega^{ab} \zeta_1 
 + \frac{ic_3}{3} B^{ab} e_a \gamma_b \zeta_1 - \frac{ic_3}{6}
\Gamma^{abc} B_{ab} e_c \zeta_1 \nonumber \\
\delta \psi &=& 2c_2 \Gamma^{ab} \eta^{ab} \phi + c_4 \Gamma^{ab}
B^{ab} \zeta_1 + ic_7 \gamma^a \pi^a \zeta_1
\end{eqnarray}

\noindent
{\bf Terms without derivatives.} The most general cubic terms without
derivatives can be written as follows:
\begin{eqnarray}
{\cal L}_{01} &=& d_1 \hat{E}_{abc} \bar{\Phi} \Gamma^{ab} \Psi f^c +
d_2 \hat{E}_{abc} \bar{\Phi} \Gamma^{abcd} \Psi f^d + id_3 
\hat{E}_{ab} \bar{\Phi} \gamma^a \psi f^b + id_4 \hat{E}_{ab}
\bar{\Psi} \gamma^a \phi f^b \nonumber  \\
 && + id_5 \hat{E}_{ab} \bar{\Phi} \Gamma^{abc} \psi f^c + id_6
\hat{E}_{ab} \bar{\Psi} \Gamma^{abc} \phi f^c + d_7 \hat{e}_a
\bar{\phi} \psi f^a + d_8 \hat{e}_a \bar{\phi} \Gamma^{ab} \psi f^b
\nonumber \\
 && + id_9 \hat{E}_{abc} \bar{\Phi} \Gamma^{abc} \Psi A +
d_{10} \hat{E}_{ab} \bar{\Phi} \Gamma^{ab} \psi A + d_{11} 
\hat{E}_{ab} \bar{\Psi} \Gamma^{ab} \phi A + id_{12} \hat{e}_a
\bar{\phi} \gamma^a \psi A \nonumber \\
 && + d_{13} \hat{E}_{ab} \bar{\Phi} \Gamma^{ab} \Psi
\sigma + id_{14} \hat{e}_a \bar{\Phi} \gamma^a \psi \sigma + id_{15}
\hat{e}_a \bar{\Psi} \gamma^a \phi \sigma + d_{16} \bar{\phi} \psi
\sigma
\end{eqnarray}
while the additional corrections for the fermions look like:
\begin{eqnarray}
\delta \Phi &=& im_1c_1 (\gamma^a \xi^a \Psi - \gamma^a f^a \zeta_2) +
\beta_2 e_a \xi^a \psi + \beta_3 \Gamma^{ab} e_a \xi_b \psi \nonumber
\\
 && + (mc_1-d_9) (\Psi \xi - A \zeta_2) - \frac{id_{13}}{6} e_a
\gamma^a \sigma \zeta_2 \\
\delta \phi &=& i\beta_4 \gamma^a \xi^a \psi + (12mc_2-d_{12}) \psi
\xi + d_{15} \sigma \zeta_2 \nonumber
\end{eqnarray}
where
$$
\beta_2 = - \frac{{\cal M} + 3m_2{}^2}{6m_2} c_1, \qquad
\beta_3 = \frac{{\cal M} + m_2{}^2}{6m_2} c_1, \qquad
\beta_4 = - \frac{(2m_1-m_2){\cal M}}{3m_1m_2} c_1
$$
and
\begin{eqnarray}
\delta \Psi &=& im_2c_1 (\gamma^a \xi^a \Phi - \gamma^a f^a \zeta_1) +
\tilde{\beta}_2 e_a \xi^a \phi + \tilde{\beta}_3 \Gamma^{ab} e_a \xi_b
\phi \nonumber \\
 && + (mc_1+d_9) (\Phi \xi - A \zeta_1) - \frac{id_{13}}{6} e_a
\gamma^a \sigma \zeta_1 \\
\delta \psi &=& i\tilde{\beta}_4 \gamma^a \xi^a \phi + (12mc_2+d_{12})
\phi \xi + d_{14} \sigma \zeta_1 \nonumber
\end{eqnarray}
where
$$
\tilde{\beta}_2 = - \frac{{\cal M} + 3m_1{}^2}{6m_1} c_1, \qquad
\tilde{\beta}_3 = \frac{{\cal M} + m_1{}^2}{6m_1} c_1, \qquad
\tilde{\beta}_4 = \frac{(m_1-2m_2){\cal M}}{3m_1m_2} c_1
$$
Here to simplify presentation we introduced a combination
$$
{\cal M} = m^2 - (m_1{}^2 + m_1m_2 + m_2{}^2)
$$
As in the both previous cases all the coefficients can be expressed in
term of the one main coefficient $c_1$, but to simplify formulas we
give here their expressions in terms of the $c_1$ and $c_2$ that are
not independent but satisfy the relation
$$
12m_1m_2c_2 = - {\cal M}c_1
$$
All other coefficients then look like:
\begin{eqnarray*}
2mc_3 &=& 3(m_1-m_2)c_1, \qquad
2mc_4 = 3m_2c_1 - 12m_1c_2 \\
2mc_5 &=& 3m_1c_1 - 12m_2c_2, \qquad
mc_6 = 4(m_1-m_2)c_2 \\
m^2c_7 &=& 2m_2{}^2c_1 + 8m_1(m_1-2m_2)c_2, \qquad
m^2c_8 = 2m_1{}^2c_1 - 8m_2(2m_1-m_2)c_2 \\
2d_1 &=& 3(m_1+m_2)c_1, \quad 
2d_2 = (m_1-m_2)c_1 \\
d_3 &=& m_2c_1 - 12m_1c_2, \quad
d_4 = m_1c_1 - 12m_2c_2 \\
d_5 &=& - m_2c_1/2 + 3m_1c_2, \quad
d_6 = - m_1c_1/2 + 3m_2c_2 \\
d_7 &=& 2(m_1+m_2)c_2, \quad
d_8 = 4(m_1-m_2)c_2 \\
md_9 &=& 2(m_1{}^2-m_2{}^2)c_1, \quad
d_{10} = d_{11} = 0, \quad
md_{12} = 8(m_1{}^2-m_2{}^2)c_2 \\
m^2d_{13} &=& - 3(m_1{}^3+m_2{}^3)c_1 + 12m_1m_2(m_1+m_2)c_2 \\
m^2d_{14} &=& - m_2(3m^2 - 2m_1m_2 - 4m_2{}^2)c_1 
- 4m_1(- 3m^2 - 2m_1{}^2 + 8m_2{}^2)c_2 \\
m^2d_{15} &=&  - m_1(3m^2 - 4m_1{}^2 - 2m_1m_2)c_1 
- 4m_2( - 3m^2 + 8m_1{}^2 - 2m_2{}^2)c_2 \\
3m^2d_{16} &=& - 8(m_1+m_1)(m^2+2(m_1-m_2)^2)c_2
\end{eqnarray*}
{\bf Massless limits} From the expressions above it follows that the
limit of massless graviton $m \to 0$ is possible for the equal masses
for the two gravitini $m_1 = m_2$ only. Again it is in agreement with
the general properties of such cubic vertices \cite{Met07b}. Indeed
the cubic vertex for the massless spin-2 and two massive spin-3/2 with
different masses does exist but requires as many as four derivatives.
From the other hand the limit $m_2 \to 0$ then one of the gravitini
becomes massless is possible for the equal masses $m = m_1$ for the
graviton and massive gravitini only in agreement with the results of
the previous section.

\noindent
{\bf Algebraic structure} Again it is instructive to consider the
commutators in the lowest non-trivial order. For the bosonic fields we
may take the commutators of the two supertransformations and obtain:
$$
[\delta_1, \delta_2] f^a = D \tilde{\xi}^a - \tilde{\eta}^{ab} e_b + m
e^a \tilde{\xi}, \qquad [\delta_1, \delta_2] A = D \tilde{\xi}
+\frac{m}{2} e_a \tilde{\xi}^a, \qquad [\delta_1, \delta_2] \sigma =
3m \tilde{\xi}
$$
\begin{equation}
\tilde{\xi}^a = 6ic_1 (\bar{\zeta}_2 \gamma^a \zeta_1), \quad
\tilde{\eta}^{ab} = 3(m_1+m_2)c_1 (\bar{\zeta}_2 \Gamma^{ab} \zeta_1),
\quad \tilde{\xi} = - \frac{3(m_1-m_2)c_1}{m} (\bar{\zeta}_2 \zeta_1)
\end{equation}
For the first gravitino we may take the commutators of the bosonic and
second supertransformations:
$$
[ \delta_B, \delta_\zeta] \Phi = (D + \frac{im_1}{2} e_a \gamma^a)
\tilde{\zeta}_1, \qquad [ \delta_B, \delta_\zeta] \phi = 3m_1
\tilde{\zeta}_1
$$
\begin{equation}
\tilde{\zeta}_1 = - \frac{c_1}{2} (\Gamma^{ab} \eta^{ab} \zeta_2)
 - im_1c_1 (\gamma^a \xi^a \zeta_2) 
- \frac{(m^2-2m_1{}^2+2m_2{}^2)c_1}{m} (\xi \zeta_2)
\end{equation}
while for the second gravitino --- the commutators of the bosonic and
first supertransformations:
$$
[ \delta_B, \delta_\zeta] \Psi = (D + \frac{im_2}{2} e_a \gamma^a)
\tilde{\zeta}_2, \qquad [ \delta_B, \delta_\zeta] \psi = 3m_2
\tilde{\zeta}_2
$$
\begin{equation}
\tilde{\zeta}_2 = - \frac{c_1}{2} (\Gamma^{ab} \eta^{ab} \zeta_1)
 - im_2c_1 (\gamma^a \xi^a \zeta_1)
- \frac{(m^2 + 2m_1{}^2-2m_2{}^2)c_1}{m} (\xi \zeta_1)
\end{equation}
It is easy to check that for the case $m_1 = m_2$ these expressions
correctly reproduce the results of Section \ref{case1}, while for the
case $m_2 = 0$ --- the results of Section \ref{case2}.

\appendix

\section{One derivative vertices}

In this Appendix we analyze all possible cubic terms for the bosonic
spin-2,1,0 and fermionic spin-3/2,1/2 components with one derivative.
Following our general formalism we consider terms that do not vanish
(or are not equivalent) on-shell and the main requirement is that all
variations with the highest number of derivatives can be compensated
by the appropriate corrections to the gauge transformations. As for
the less derivative variations we take them into account in the main
text together with the variations of the terms without derivatives.

\subsection{Subvertex $2-3/2-3/2$}

In this case the only possibility is:
\begin{equation}
{\cal L}_{1a} = ic_1 \hat{E}_{abc} \Omega^{ab} \bar{\Phi} \gamma^c
\Phi + ic_2 \hat{E}_{abcd} \bar{\cal F} \Gamma^{abc} \Phi f^d
\end{equation}
Let us consider $\zeta$-transformations first. They produce the
following variations:
\begin{eqnarray*}
\delta_\zeta {\cal L}_{1a} &=& - 2ic_1 \hat{E}_{abc} {\cal R}^{ab}
\bar{\Phi} \gamma^c \zeta
 + 2ic_1 \hat{E}_{abc} \Omega^{ab} \bar{\cal F} \gamma^c \zeta 
 - 3ic_2 \hat{E}_{abc} \Omega^{ad} \bar{\cal F} \Gamma^{bcd} \zeta \\
 && + 4m_1c_1 \hat{E}_{ab} \Omega^{ac} \bar{\Phi} \Gamma^{bc} \zeta
  + \frac{4im_1c_1}{3} \hat{e}_a \Omega^{ab} \bar{\phi} \gamma^b \zeta
 + \frac{im_1c_1}{3} \hat{e}_a \Omega^{bc} \bar{\phi} \Gamma^{abc}
\zeta \\
 && + m_1c_2 \hat{E}_{abc} \bar{\cal F} \Gamma^{abcd} \zeta f^d 
 - im_1c_2 \hat{E}_{ab} \bar{\cal C} \gamma^a \zeta f^b 
- \frac{im_1c_2}{2} \hat{E}_{ab} \bar{\cal C} \Gamma^{abc} \zeta f^c 
\\
 && - 4imc_1 \hat{e}_a B^{ab} \bar{\Phi} \gamma^b \zeta 
 - imc_2 \hat{E}_{abc} \bar{\cal F} \Gamma^{abc} \zeta A \\
 && - 4im^2c_1 \hat{E}_{ab} \bar{\Phi} \gamma^a \zeta f^b 
+ 4im^2c_1 \hat{e}_a \bar{\Phi} \gamma^a \zeta \sigma
\end{eqnarray*}
To compensate the terms with the highest number of derivatives (the
first line) we introduce the following corrections:
\begin{equation}
\delta_1 f^a = i\alpha_1 \bar{\Phi} \gamma^a \zeta, \qquad
\delta_1 \Phi = \alpha_2 \Gamma^{ab} \Omega^{ab} \zeta
\end{equation}
Their contribution looks like:
$$
\delta_1 {\cal L}_0 = - \frac{i\alpha_1}{2} \hat{E}_{abc} 
{\cal R}^{ab} \bar{\Phi} \gamma^c \zeta + 6i\alpha_2 \hat{E}_{abc}
\bar{\cal F} ( \Omega^{ab} \gamma^c - \Gamma^{abe} \Omega^{ce}) \zeta 
$$
Thus we have to put:
$$
\alpha_1 = - 4c_1, \qquad 6\alpha_2 = - 2c_1 = - 3c_2
$$
Let us turn to the $\eta^{ab}$-transformations. They produce:
\begin{eqnarray*}
\delta_\eta {\cal L}_{1a} &=& - 2ic_1 \hat{E}_{abc} \eta^{ab}
\bar{\cal F} \gamma^c \Phi 
 + 3ic_2 \hat{E}_{abc} \eta^{ad} \bar{\cal F} \Gamma^{bcd} \Phi \\
 && + 2M_1c_1 \hat{E}_{ab} \eta^{ac} \bar{\Phi} \Gamma^{bc} \Phi 
 + \frac{4im_1c_1}{3} \hat{e}_a \eta^{ab} \bar{\Phi} \gamma^b \phi 
- \frac{im_1c_1}{3} \hat{e}_a \eta^{bc} \bar{\Phi} \Gamma^{abc} \phi
\end{eqnarray*}
To compensate for the terms in the first line we introduce:
\begin{equation}
\delta_1 \Phi = \alpha_3 \Gamma^{ef} \eta^{ef} \Phi
\end{equation}
and this gives us
$$
\delta_1 {\cal L}_0 = 6i\alpha_3 \hat{E}_{abc} \bar{\cal F}
( \eta^{ab} \gamma^c - \eta^{ad} \Gamma^{bcd}) \Phi
$$
So we obtain (in agreement with the previous relation on $c_{1,2}$):
$$
6\alpha_3 = 2c_1 = 3c_2
$$

\subsection{Subvertex $2-1/2-1/2$}

In this case we consider the terms:
\begin{equation}
{\cal L}_{1b} = ic_3 \hat{e}_a \Omega^{bc} \bar{\phi} \Gamma^{abc}
\phi + ic_4 \hat{E}_{ab} \bar{\cal C} \gamma^a \phi f^b
\end{equation}
Note that there is one more possible term $\hat{E}_{ab} \bar{\cal C}
\Gamma^{abc} \phi f^c$ but (up to the terms without derivatives) it is
equivalent to the term with coefficient $c_3$.

The only transformations that produce variations with the highest
number of derivatives are $\eta^{ab}$-transformations:
\begin{eqnarray*}
\delta_\eta {\cal L}_{1b} &=& ic_4 \hat{e}_a \eta^{ab} \bar{\cal C}
\gamma^b \phi 
 - 2ic_3 \hat{e}_a \eta^{bc} \bar{\cal C} \Gamma^{abc} \phi \\
 && - 6im_1c_3 \hat{e}_a \eta^{bc} \bar{\Phi} \Gamma^{abc} \phi 
\end{eqnarray*}
To compensate for the terms in the first line we introduce
\begin{equation}
\delta_1 \phi = \alpha_4 \Gamma^{ef} \eta^{ef} \phi
\end{equation}
which gives:
$$
\delta_1 {\cal L}_0 = - i\alpha_4 \hat{e}_a \bar{\cal C} ( 2 \eta^{ab}
\gamma^c + \Gamma^{abc} \eta^{bc}) \phi 
$$
So we obtain:
$$
2\alpha_4 = - 4c_3 = c_4
$$

\subsection{Subvertex $1-3/2-1/2$}

For this case we consider:
\begin{equation}
{\cal L}_{1c} = ic_5 \hat{e}_a B^{ab} \bar{\Phi} \gamma^b \phi + ic_6
\hat{e}_a B^{bc} \bar{\Phi} \Gamma^{abc} \phi
\end{equation}
Note that there are two more possible terms: 
$\hat{E}_{abc} \bar{\cal F} \Gamma^{abc}\phi A$ and 
$\hat{E}_{abc} \bar{\Phi} \Gamma^{abc} {\cal C} A$. However,
one their combination is equivalent (up to the terms without
derivatives) to the term with coefficient $c_6$. Besides, there
exists a field redefinition $\Phi \Rightarrow \Phi + \kappa_1 A \phi$.

The only transformations we have to consider here are 
$\zeta$-transformations which produce the following variations:
\begin{eqnarray*}
\delta_\zeta {\cal L}_{1c} &=& ic_5 \hat{e}_a {\cal B}^{ab} \bar{\phi}
\gamma^b \zeta  
 + ic_5 \hat{e}_a B^{ab} \bar{\cal C} \gamma^b \zeta
- ic_6 \hat{e}_a B^{bc} \bar{\cal C} \Gamma^{abc} \zeta \\
 && - imc_5 \hat{e}_a \Omega^{ab} \bar{\phi} \gamma^b \zeta 
 + imc_6 \hat{e}_a \Omega^{bc} \bar{\phi} \Gamma^{abc} \zeta \\
 && + 6im_1c_5 \hat{e}_a B^{ab} \bar{\Phi} \gamma^b \zeta  
- m_1c_5 B^{ab} \bar{\phi} \Gamma^{ab} \zeta   \\
 && + imc_5 \bar{\phi} \gamma^a \zeta \pi^a 
\end{eqnarray*}
To compensate for the terms in the first line we introduce:
\begin{equation}
\delta_1 A = i\alpha_5 \bar{\phi} e_e \gamma^e \zeta, \qquad
\delta_1 \phi = \alpha_6 \Gamma^{ef} B^{ef} \zeta
\end{equation}
and this gives us
$$
\delta_1 {\cal L}_0 = 2i\alpha_5 \hat{e}_a {\cal B}^{ab} \bar{\phi}
\gamma^b \zeta - i\alpha_6 \hat{e}_a \bar{\cal C} ( 2 B^{ab} \gamma^b
+ B^{bc} \Gamma^{abc}) \zeta
$$
Thus we have to put:
$$
2\alpha_5 = - c_5, \qquad 2\alpha_6 = c_5 = - 2c_6
$$

\subsection{Subvertex $0-3/2-1/2$}

For this case we choose:
\begin{equation}
{\cal L}_{1d} = c_7 \hat{e}_a \bar{\Phi} \phi \pi^a + c_8 \hat{e}_a
\bar{\Phi} \Gamma^{ab} \phi \pi^b
\end{equation}
Note that there are two more possible terms: 
$\hat{E}_{ab} \bar{\cal F} \Gamma^{ab} \phi \sigma$ and 
$\hat{E}_{ab} \bar{\Phi} \Gamma^{ab} {\cal C} \sigma$.
However, one their combination is equivalent (up to the terms without
derivatives) to the term with coefficient  $c_8$. Besides, there
exists a field redefinition  $\Phi \Rightarrow \Phi + \kappa_2 \sigma
e^a \gamma^a \phi$.

In this case we have to consider $\zeta$-transformations only:
\begin{eqnarray*}
\delta_\zeta {\cal L}_{1d} &=& - c_7 \hat{e}_a \bar{\phi} \zeta \Pi^a 
 - c_7 \hat{e}_a \bar{\cal C} \zeta \pi^a 
 + c_8 \hat{e}_a \bar{\cal C} \Gamma^{ab} \zeta \pi^b \\
 && - \frac{3mc_8}{2} B^{ab} \bar{\phi} \Gamma^{ab} \zeta 
 + 6m_1c_8 \hat{e}_a \bar{\Phi} \Gamma^{ab} \zeta \pi^b 
 + 3im_1c_8 \bar{\phi} \gamma^a \zeta \pi^a     \\
 && + \frac{3m^2c_7}{2} \hat{e}_a \bar{\phi} \zeta f^a
 - \frac{3m^2c_8}{2} \hat{e}_a \bar{\phi} \Gamma^{ab} \zeta f^b 
 - 2m^2c_7 \bar{\phi} \zeta \sigma
\end{eqnarray*}
To compensate for the terms in the first line we introduce:
\begin{equation}
\delta_1 \sigma = \alpha_7 \bar{\phi} \zeta, \qquad
\delta_1 \phi = i\alpha_8 \gamma^b \pi^b \zeta
\end{equation}
Taking into account their contributions:
$$
\delta_1 {\cal L}_0 = - \frac{2\alpha_7}{3} \hat{e} \bar{\phi} \zeta
\Pi^a + \alpha_8 \hat{e}_a \bar{\cal C} ( \pi^a + \Gamma^{ab} \pi^b)
\zeta
$$
we obtain:
$$
\alpha_7 = - \frac{3c_7}{2}, \qquad \alpha_8 = c_7 = - c_8
$$


\begin{thebibliography}{10}

\bibitem{RGT10}
Claudia de~Rham, Gregory Gabadadze, Andrew~J. Tolley
{\it "Resummation of Massive Gravity",}
Phys. Rev. Lett. {\bf 106} (2011) 231101, arXiv:1011.1232.

\bibitem{RGT11}
Claudia de~Rham, Gregory Gabadadze, Andrew Tolley
{\it "Ghost free Massive Gravity in the Stueckelberg language",}
Phys. Lett. {\bf B711} (2012) 190, arXiv:1107.3820.

\bibitem{BD72}
D.~G. Boulware, S.~Deser
{\it "Can gravitation have a finite range?",}
Phys. Rev. {\bf D6} (1972) 3368.

\bibitem{HRS11}
S.~F. Hassan, Rachel~A. Rosen, Angnis Schmidt-May
{\it "Ghost-free Massive Gravity with a General Reference Metric",}
JHEP {\bf 1202} (2012) 026, arXiv:1109.3230.

\bibitem{HR11}
S.~F. Hassan, Rachel~A. Rosen
{\it "Bimetric Gravity from Ghost-free Massive Gravity",}
JHEP {\bf 1202} (2012) 126, arXiv:1109.3515.

\bibitem{HR11b}
S.~F. Hassan, Rachel~A. Rosen
{\it "Confirmation of the Secondary Constraint and Absence of Ghost in
Massive Gravity and Bimetric Gravity",}
JHEP {\bf 1204} (2012) 123, arXiv:1111.2070.

\bibitem{Mal13}
Ola Malaeb
{\it "Massive Gravity with N=1 local Supersymmetry",}
Eur. Phys. J. {\bf C73} (2013) 2549, arXiv:1302.5092.

\bibitem{Mal13a}
Ola Malaeb
{\it "Supersymmetrizing Massive Gravity",}
Phys. Rev. {\bf D88} (2013) 025002, arXiv:1303.3580.

\bibitem{MFG16}
F.~Del Monte, D.~Francia, P.~A. Grassi
{\it "Multimetric Supergravities",}
JHEP {\bf 09} (2016) 064, arXiv:1605.06793.

\bibitem{OT16}
Nicholas~A. Ondo, Andrew~J. Tolley
{\it "Deconstructing Supergravity, I: Massive Supermultiplets",}
  arXiv:1612.08752.

\bibitem{Met07b}
R.~R. Metsaev
{\it "Cubic interaction vertices for fermionic and bosonic arbitrary
spin fields",}
Nucl. Phys. {\bf B859} (2012) 13, arXiv:0712.3526.

\bibitem{Zin09}
Yu.~M. Zinoviev
{\it "On massive spin 2 electromagnetic interactions",}
Nucl. Phys. {\bf B821} (2009) 431-451, arXiv:0901.3462.

\bibitem{BSZ11}
Nicolas Boulanger, E.~D. Skvortsov, Yu.~M. Zinoviev
{\it "Gravitational cubic interactions for a simple mixed-symmetry
gauge field in AdS and flat backgrounds",}
J. Phys. {\bf A44} (2011) 415403, arXiv:1107.1872.

\bibitem{Zin01}
Yu.~M. Zinoviev
{\it "On Massive High Spin Particles in (A)dS",} arXiv:hep-th/0108192.

\bibitem{Met06}
R.~R. Metsaev
{\it "Gauge invariant formulation of massive totally symmetric
fermionic fields in (A)dS space",}
Phys. Lett. {\bf B643} (2006) 205-212, arXiv:hep-th/0609029.

\bibitem{Zin08b}
Yu.~M. Zinoviev
{\it "Frame-like gauge invariant formulation for massive high spin
particles",}
Nucl. Phys. {\bf B808} (2009) 185, arXiv:0808.1778.

\bibitem{PV10}
D.~S. Ponomarev, M.~A. Vasiliev
{\it "Frame-Like Action and Unfolded Formulation for Massive
Higher-Spin Fields",}
Nucl. Phys. {\bf B839} (2010) 466, arXiv:1001.0062.

\bibitem{BSZ14}
I.~L. Buchbinder, T.~V. Snegirev, Yu.~M. Zinoviev
{\it "Formalism of gauge invariant curvatures and constructing the
cubic vertices for massive spin-3/2 field in $AdS_4$ space",}
Eur. Phys. J. C {\bf 74} (2014) 3153, arXiv:1405.7781.

\end{thebibliography}
\end{document}